\documentclass[12pt]{article}
\usepackage{epsfig}
\def\bea{\begin{eqnarray}}
\def\eea{\end{eqnarray}}
\def\bq{\begin{quote}}
\def\eq{\end{quote}}

\def\gappeq{\mathrel{\rlap
{\raise.5ex\hbox{$>$}}
{\lower.5ex\hbox{$\sim$}}}}
\def\lappeq{\mathrel{\rlap{\raise.5ex\hbox{$<$}}
{\lower.5ex\hbox{$\sim$}}}}
\def\simlt{\stackrel{<}{{}_\sim}}
\def\simgt{\stackrel{>}{{}_\sim}}
\newcommand{\ynu}[1]{\mathbf{Y}_{\nu}^{#1}}

\newcommand{\beq}{\begin{equation}}
\newcommand{\eeq}{\end{equation}}
\newcommand{\OO}[1]{\mathbf{\Omega}_{#1}} 
\newcommand{\OP}[1]{\mathbf{\Omega'}_{#1}} 
\newcommand{\YY}[1]{\mathbf{\tilde{Y}}_\nu^{#1}} 
\newcommand{\Un}[1]{\mathbf{U}_\nu^{#1}} 
\newcommand{\Ur}[1]{\mathbf{U}_N^{#1}}

\def\ch{\mathrm{cosh}\,}
\def\sh{\mathrm{sinh}\,}
\def\th{\mathrm{tanh}\,}

\begin{document}
\pagestyle{empty}
\begin{flushright}
{\bf IFT-04/05\\
\bf hep-ph/0401219\\
\bf \today}
\end{flushright}
\vspace*{5mm}
\begin{center}
{\bf Degenerate minimal see-saw and leptogenesis.}
\\
\vspace*{1cm} 
Krzysztof Turzy\'nski
\\
Institute of Theoretical Physics, Warsaw University, \\
Ho\.za 69, 00-681 Warsaw, Poland 
\vskip 1.0cm

\vspace*{1.7cm} 
{\bf Abstract} 
\end{center}
\vspace*{5mm}
\noindent
{\small We analyze the most general version of the supersymmetric minimal see-saw model with only two right-chiral neutrinos which are degenerate in masses at the scale of Grand Unification. We study the renormalization effects that give rise to non-zero CP asymmetries in the decays of these neutrinos and find that their resonant enhancement due to small mass splitting is partly compensated by other RG effects in the running of the neutrino Yukawa couplings. In spite of this compensation, the CP asymmetries can be large enough for successful thermal leptogenesis. Moreover, they depend very weakly on the right-chiral neutrino masses and the resulting leptogenesis can be successful for very low reheating temerature, thereby allowing to overcome the gravitino problem. 
}
\vspace*{1.0cm}
\date{\today} 


\vspace*{0.2cm}

\vfill\eject
\newpage

\setcounter{page}{1}
\pagestyle{plain}

The minimal version of the see-saw mechanism \cite{GERASL} can be obtained by adding two very heavy gauge singlets $N^1$ and $N^2$ to the spectrum of the Standard Model \cite{Frampton:2002qc,Kuchimanchi:2002yu}. Their interactions can be described by the following potential:
\beq
\label{potential}
\Delta\mathcal{L} = -\epsilon_{ij}H_i N^K \ynu{KA} \ell^A_j - \frac{1}{2}\mathbf{M}^{KL} N^K N^L + H.c.
\eeq
where $H$ and $\ell^A$ are Higgs and lepton doublets, $N^K$, $K=1,2$ denote the right-chiral neutrinos, $\mathbf{M}^{KL}$ is their Majorana mass matrix and $\ynu{KA}$ is the $2\times 3$ matrix of the neutrino Yukawa couplings. 
Integrating out the fields $N^1$ and $N^2$ provides (after the electroweak symmetry breaking in the low-energy
effective theory) the left-chiral neutrinos with small Majorana masses:
\beq
\mathbf{m}_\nu = -\frac{\langle H\rangle^2}{2} \ynu{T} \mathbf{M}^{-1} \ynu{}
\eeq
where $\mathbf{m}_\nu$ has only two non-zero eigenvalues $m_{\nu_2}$ and $m_{\nu_3}$, as the rank of $\mathbf{m}_\nu$ cannot exceed that of $\mathbf{M}$. Their values are $m_{\nu_2}=\sqrt{\Delta m_\mathrm{sol}^2}$, $m_{\nu_3}\approx \sqrt{\Delta m_\mathrm{atm}^2}$ (normal hierarchy) or $m_{\nu_2},\,\,m_{\nu_3}\approx\sqrt{\Delta m_\mathrm{atm}^2}$, $m_{\nu_3}^2-m_{\nu_2}^2=\Delta m_\mathrm{sol}^2$ (inverse hierarchy), which accounts for the solar and atmospheric oscillations with differences of the masses squared $\Delta m_\mathrm{sol}^2\approx 7\times 10^{-5}\,\mathrm{eV}$ and $\Delta m_\mathrm{atm}^2\approx 2\times 10^{-3}\,\mathrm{eV}$. 

It is known that the CP violating decays of the right-chiral neutrinos can produce lepton asymmetry, which is subsequently transformed into baryon asymmetry via sphaleron transitions \cite{Fukugita:1986hr}. This possibility of explaining the measured baryon asymmetry of the Universe \cite{Spergel:2003cb} has been investigated in many papers with either minimal \cite{Frampton:2002qc,Barger:2003gt,Ibarra:2003up} or non-minimal see-saw models (see e.g. \cite{Buchmuller:2003gz}).

It has been shown \cite{Chankowski:2003rr} that with the minimal see-saw mechanism, the wash-out processes are generically very efficient and one needs the reheating temperature after inflation $T_{RH}\simgt 2\times 10^{11}\,\mathrm{GeV}$ in order to generate sufficient lepton asymmetry in the decays of the lightest right-chiral neutrino\footnote{this bound is derived by assuming that $T_{RH}$ is larger than the relevant right-chiral neutrino mass, not at least 10 times larger, as in \cite{Chankowski:2003rr}}. This creates a problem, because in supersymmetric theories, overpoduction of gravitinos that destroys nucleosynthesis \cite{gravitino} may occur already for $T_{RH}\simgt 10^7\,\mathrm{GeV}$ \cite{g_bound}. 

One possible way of circumventing this problem is to assume that the two right-chiral neutrinos are almost degenerate in masses, which enhances the CP asymmetries in the decays of both right-chiral neutrinos contributing to the lepton asymmetry \cite{Flanz:1994yx} and allows lowering the reheating temperature \cite{Ellis:2002eh,Chankowski:2003rr}. The natural idea that such a tiny splitting of the masses of the right-chiral neutrinos might result from the renormalization group (RG) evolution of the neutrino parameters with exact mass degeneracy at some higher (presumably the GUT) scale has been recently investigated in \cite{GonzalezFelipe:2003fi} in the framework of a model with one texture zero in the neutrino Yukawa matrix. Furthermore, in \cite{GonzalezFelipe:2003fi} the element $\Un{13}$  of the neutrino mixing matrix was required to be very small compared to the maximal value allowed by experiment. The purpose of this note is to extend these results to general minimal see-saw models with degenerate right-chiral neutrinos and to point out some important features of the RG analysis which have not been taken into account before. Motivated by the gravitino problem, we perform the calculations in the MSSM.

In the full theory at high energies, one can remove the unphysical parameters by a unitary transformation $N^K\to \Ur{KL}N^L$ which diagonalizes the mass matrix $\mathbf{M}$, i.e. $\Ur{T}\mathbf{M}\Ur{}=\textrm{diag}(M_1,M_2)$. Similarly, the lepton doublets are transformed to the basis in which the Yukawa matrix of the charged leptons is diagonal, $\ell^A\to \mathbf{U}_\ell^{AB}\ell^B$.  Then the neutrino Yukawa couplings also change, i.e. $\ynu{}\to \Ur{T}\ynu{}\mathbf{U}_\ell$. This basis is convenient, since it directly corresponds to the interactions of the separate right-chiral neutrinos. The RG equations for $M_i$ and $\ynu{}$ written in this basis can be obtained by using the methods outlined in \cite{Babu:im,Chankowski:2001mx}:
\begin{eqnarray}
\label{RGEM}
\frac{d\,}{dt} M_K &=& 4\left( \ynu{}\ynu{\dagger}\right)_{KK} M_K \\
\frac{d\,}{dt} \ynu{KA} &=& K_Y \ynu{KA} + y_{e_A}^2 \ynu{KA}+ 3\left( \ynu{}\ynu{\dagger}\right)_{KL}\ynu{LA} - \nonumber\\
& & -2(1-\delta_{LK})\left(\frac{M_L+M_K}{M_L-M_K}\mathsf{Re}\left( \ynu{}\ynu{\dagger}\right)_{KL}+i\frac{M_L-M_K}{M_L+M_K}\mathsf{Im}\left( \ynu{}\ynu{\dagger}\right)_{KL} \right) \ynu{LA} + \nonumber\\
\label{RGEyukawa}
& & +(1-\delta_{AB})\frac{y_{e_B}^2+y_{e_A}^2}{y_{e_B}^2-y_{e_A}^2}\left( \ynu{\dagger}\ynu{}\right)_{AB} \ynu{KB}
\end{eqnarray}
where $K_Y=-3g_2^2-\frac{3}{5}g_1^2+3\sum_B y_{u_B}^2+\mathsf{Tr}(\ynu{\dagger}\ynu{})$, $y_{u_A}$ and $y_{e_A}$ are the Yukawa couplings of the up-type quarks and the charged leptons, respectively, and $t=\frac{1}{16\pi^2}\ln(\mu/M_X)$. 

The reduction of the number of parameters in the neutrino sector with respect to the conventional see-saw mechanism with three right-chiral neutrinos does not completely remove the ambiguity in extracting the Yukawa couplings $\ynu{KA}$ from the low energy data and the masses of the right-chiral neutrinos. Rather one has \cite{Casas:2001sr}:
\beq
\ynu{KA} = \frac{\sqrt{M_Km_{\nu_a}}}{\langle H\rangle} \OO{Ka} \Un{Aa*}
\eeq
where $\Un{}$ is the neutrino mixing matrix and $\OO{}$ is an arbitrary $2\times 3$ complex matrix satisfying the condition $\OO{}^T\OO{}=\mathrm{diag}(0,1,1)$. The matrix $\OO{}$ can be conveniently parametrized with a complex 'angle' $\omega$ \cite{Ibarra:2003up}:
\beq
\label{omegapar}
\OO{} = \left( \begin{array}{ccc} 0 & \cos\omega & \sin\omega \\ 0 & -\sin\omega & \cos\omega \end{array}\right)
\eeq
Denoting the non-zero $2\times 2$ submatrix in (\ref{omegapar}) by $\OP{}$ and substituting $\omega=\alpha-i\beta$, we obtain:
\beq
\label{oprim}
\OP{} = \left( \begin{array}{cc} \cos\alpha & \sin\alpha \\ -\sin\alpha & \cos\alpha \end{array}\right)\left( \begin{array}{cc} \ch\beta & -i\,\sh\beta \\ i\,\sh\beta & \ch\beta \end{array}\right)
\eeq

If $M_1=M_2=M$ at a certain scale $M_X$, there seems to be a freedom of rotation $N^K\to \mathbf{R}^{KL}N^L$, $\mathbf{R}^T\mathbf{R}=\mathbf{1}$ which does not affect the diagonal Majorana mass matrix of the right-chiral neutrinos, but rotates the Yukawa couplings:
\beq
\label{rotation0}
\ynu{}\to \YY{}\equiv\left( \begin{array}{cc} \cos\alpha' & \sin\alpha' \\ -\sin\alpha' & \cos\alpha' \end{array}\right)\ynu{}
\eeq
However, this apparent freedom must be utilized to assure that the RG equation (\ref{RGEyukawa}) is non-singular
\beq
\label{rotation}
\mathsf{Re}\left(\YY{}\YY{\dagger}\right)_{12} = 0
\eeq
i.e. quantum corrections unambiguously choose the basis in which the right-chiral neutrinos are mass eigenstates and their mixing matrix is a continuous function of the renormalization scale \cite{Chankowski:2001mx}. The rotation (\ref{rotation0}) can be absorbed into the definition of $\OP{}$ in (\ref{oprim}) and replaced with a constraint on $\OP{}$:
\beq
\label{consomega}
0 = \mathsf{Re}\left(\YY{}\YY{\dagger}\right)_{12}=\frac{Mm_{\nu_3}}{\langle H\rangle^2} \mathsf{Re}\left(\OP{}\mathrm{diag}(0,\rho,1)\OP{}^\dagger\right)_{12} = \frac{Mm_{\nu_3}}{\langle H\rangle^2}(1-\rho)\cos\alpha\sin\alpha
\eeq
where $\rho=m_{\nu_2}/m_{\nu_3}$. There are two physically equivalent solutions of (\ref{consomega}), namely $\cos\alpha=0$ or $\sin\alpha=0$. The rotation (\ref{rotation0}) leaves $\mathsf{Im}\left(\YY{}\YY{\dagger}\right)_{12}$ intact, since:
\beq
\label{imyy}
\mathsf{Im}\left(\YY{}\YY{\dagger}\right)_{12}=-\frac{Mm_{\nu_3}}{\langle H\rangle^2}(1+\rho)\,\ch\beta\,\sh\beta
\eeq
Of course, the constraint (\ref{rotation}) must be satisfied at the scale at which the two right-chiral neutrinos have exactly equal tree level masses. At other scales, small values of $\delta_N\equiv 1-M_1/M_2$ and $\mathsf{Re}\left(\YY{}\YY{\dagger}\right)_{12}$ are generated radiatively. The solutions of the RG equations (\ref{RGEM}) and (\ref{RGEyukawa}) in the leading logarithm approximation are:
\begin{eqnarray}
\label{soldelta}
\delta_N &\approx & \pm\frac{4Mm_{\nu_3}(1-\rho)}{\langle H\rangle^2}\Delta t \\
\label{solre}
\mathsf{Re}\left(\YY{}\YY{\dagger}\right)_{12} &\approx & \pm\sqrt{\rho} \frac{Mm_{\nu_3}}{\langle H\rangle^2} \mathsf{Re}\left(\Un{32*}\Un{33}\right) y_\tau^2 \Delta t
\end{eqnarray}
where $\Delta t=\frac{1}{16\pi^2}\ln (M_X/M)$. 

A nonzero value of $\delta_N$ leads to nonvanishing CP asymmetries at the scale $M$ at which the right-chiral neutrinos decay. For small mass splittings these asymmetries are given by \cite{Pilaftsis:1998pd}:
\beq
\epsilon_i \approx - \frac{\mathsf{Im}\left(\left(\YY{}\YY{\dagger}\right)_{12}^2\right)}{\left(\YY{}\YY{\dagger}\right)_{11}\left(\YY{}\YY{\dagger}\right)_{22}} \times \frac{8\pi\delta_N\left(\YY{}\YY{\dagger}\right)_{ii}}{(8\pi\delta_N)^2+\left(\YY{}\YY{\dagger}\right)_{ii}^2} 
\eeq
In the two different regimes this formula can be further simplified:
\begin{eqnarray}
\label{cpassdefNH}
\epsilon_i \approx - \frac{\mathsf{Im}\left(\left(\YY{}\YY{\dagger}\right)_{12}^2\right)}{8\pi\delta_N \left(\YY{}\YY{\dagger}\right)_{ii}} \propto \frac{\mathsf{Re}\left(\YY{}\YY{\dagger}\right)_{12}}{\delta_N} & \textrm{for} & \delta_N\gg \left(\YY{}\YY{\dagger}\right)_{ii}/8\pi \\
\label{cpassdefIH}
\epsilon_i = - \frac{8\pi\,\mathsf{Im}\left(\left(\YY{}\YY{\dagger}\right)_{12}^2\right)\,\delta_N}{\left(\YY{}\YY{\dagger}\right)_{ii}\left(\YY{}\YY{\dagger}\right)_{11}\left(\YY{}\YY{\dagger}\right)_{22}} & \textrm{for} & \delta_N\ll \left(\YY{}\YY{\dagger}\right)_{ii}/8\pi
\end{eqnarray}
The factor $(1-\rho)$ is larger by two orders of magnitude for the normal hierarchy of the light neutrino masses compared to the inverse hierarchy. Consequently, as follows from (\ref{soldelta}), the values of $\delta_N$ generated radiatively are much smaller in the former case. It is easy to check that for $M_X\simgt 10M$ the formula (\ref{cpassdefNH}) applies to the normal hierarchy of the light neutrino masses, while (\ref{cpassdefIH}) corresponds to the inverse hierarchy. Substituting the results (\ref{imyy}), (\ref{soldelta}) and (\ref{solre}) to (\ref{cpassdefNH}) we obtain:
\begin{eqnarray}
\label{soleps1}
\epsilon_1 &\approx& \frac{\sqrt{\rho}(1+\rho)\th\beta}{16\pi (1-\rho)(\rho+\mathrm{tanh}^2\beta)} \mathsf{Re}\left(\Un{32*}\Un{33}\right) y_\tau^2 \\
\label{soleps2}
\epsilon_2 &\approx& \frac{\sqrt{\rho}(1+\rho)\th\beta}{16\pi (1-\rho)(1+\rho\,\mathrm{tanh}^2\beta)} \mathsf{Re}\left(\Un{32*}\Un{33}\right) y_\tau^2
\end{eqnarray}
for $\sin\alpha=0$. If $\cos\alpha=0$, we should interchange $\epsilon_1$ and $\epsilon_2$ in (\ref{soleps1}) and (\ref{soleps2}). As it is clear from (\ref{cpassdefNH}) in the regime $\delta_N\gg\left(\YY{}\YY{\dagger}\right)_{ii}/8\pi$ the dependence on $\Delta t$ cancels out. As a result, in the case of normal hierarchy the asymmetries $\epsilon_i$ depend on the right-chiral neutrino masses
only very weakly throught the dependence of $y_\tau^2$ on the renormalization scale. Since $y_\tau^2$ is a monotonic function of the renormalization scale, we shall use $y_\tau^2 = (y_\tau^2(M_X)+y_\tau^2(M))/2$.

Similarly, substituting (\ref{imyy}), (\ref{soldelta}) and (\ref{solre}) to (\ref{cpassdefIH}), we get:
\begin{eqnarray}
\label{soleps1ih}
\epsilon_1 &=& \frac{\sqrt{\rho}(1-\rho^2)\th\beta}{4\pi^3(\rho+\mathrm{tanh}^2\beta)(1+\rho\,\mathrm{tanh}^2\beta)^2} \mathsf{Re}\left(\Un{32*}\Un{33}\right) y_\tau^2 \ln^2\frac{M_X}{M} \\
\label{soleps2ih}
\epsilon_2 &=& \frac{\sqrt{\rho}(1-\rho^2)\th\beta}{4\pi^3(1+\rho\,\mathrm{tanh}^2\beta)(\rho+\mathrm{tanh}^2\beta)^2} \mathsf{Re}\left(\Un{32*}\Un{33}\right) y_\tau^2 \ln^2\frac{M_X}{M}
\end{eqnarray}
Since $\rho\approx 1$ for the inverse hierarchy, $\epsilon_1\approx\epsilon_2$. Note that the dependence on $M$ is much stronger for the inverse hierarchy than for the normal hierarchy due to $\ln^2(M_X/M)$ factors in (\ref{soleps1ih}) and (\ref{soleps2ih}).

\begin{figure}
\includegraphics*[height=8cm]{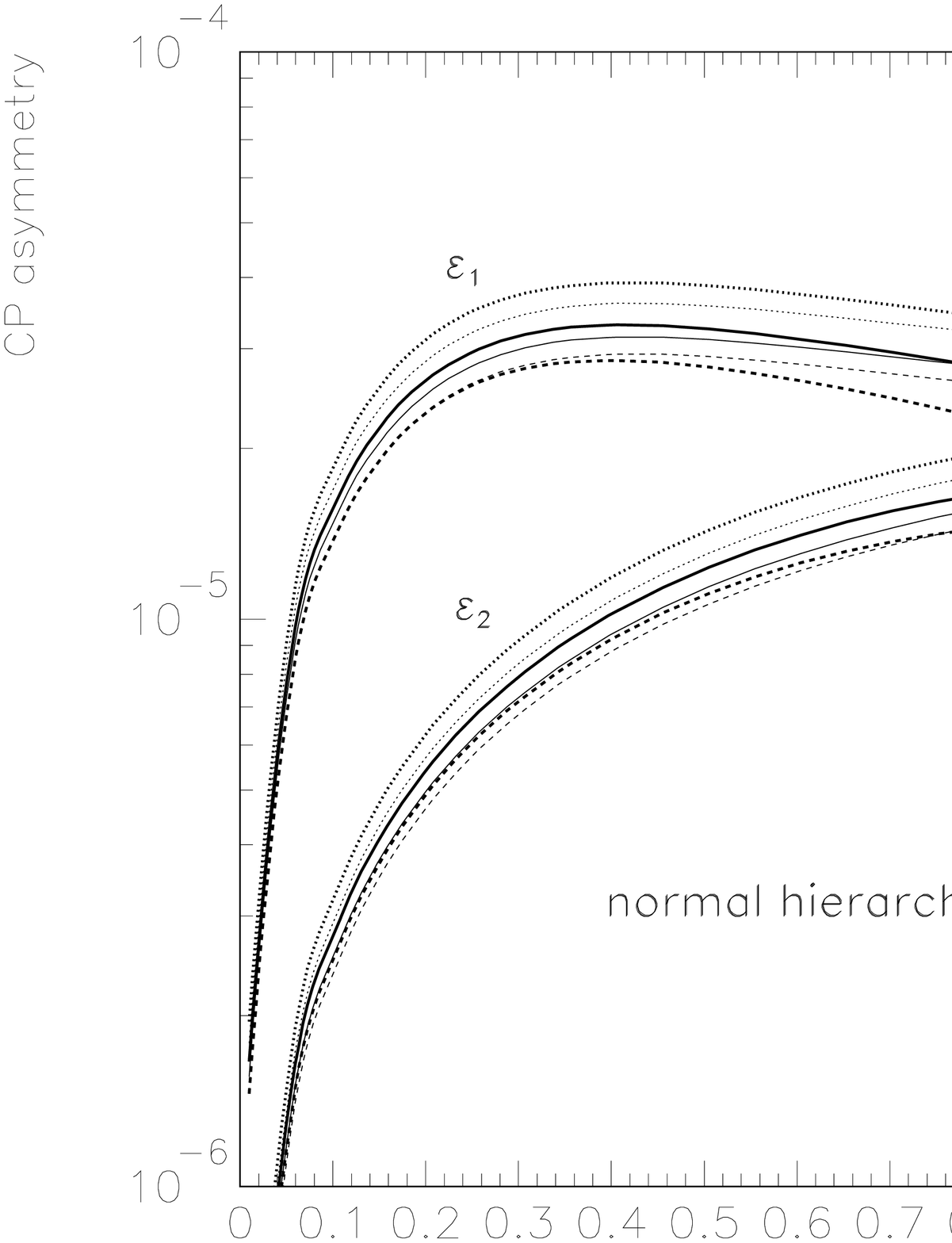}
\hspace{1cm}
\includegraphics*[height=8cm]{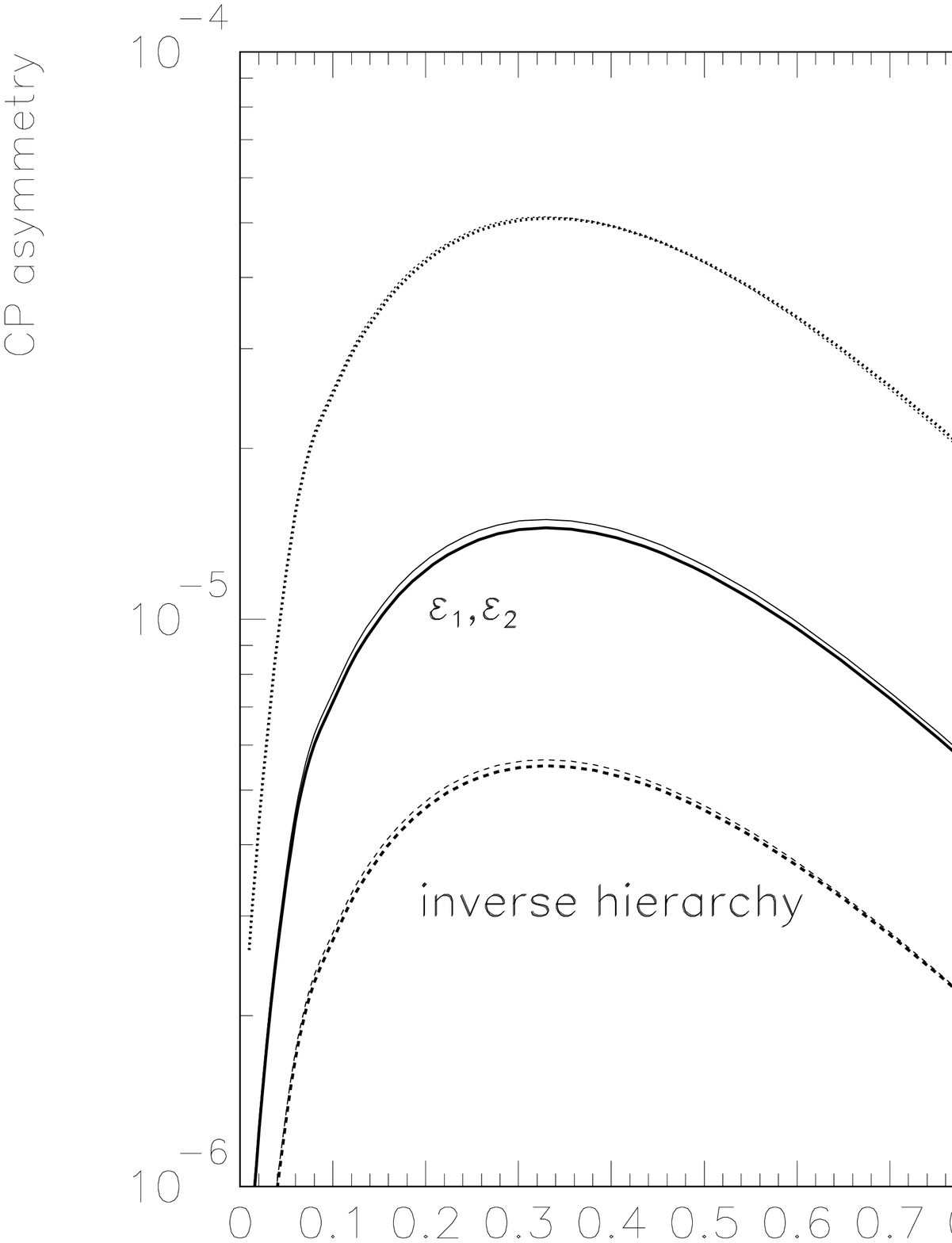}
\caption{CP asymmetries $\epsilon_1$ and $\epsilon_2$ as functions of $\beta$ for normal and inverse hierarchy of neutrino masses. Dashed, solid and dotted lines correspond to $M=10^{13},10^{11},10^7\,\mathrm{GeV}$, respectively. Thick lines correspond to numerical integration of the full RG equations and thin lines to the analytic formulae (\ref{soleps1})--(\ref{soleps2}) and (\ref{soleps1ih})--(\ref{soleps2ih}). \label{fi1}}
\end{figure}

The results of the numerical integration of the full RG equations and the approximate analytic formulae (\ref{soleps1})--(\ref{soleps2}) and (\ref{soleps1ih})--(\ref{soleps2ih}) are presented in Figure \ref{fi1} as functions of $\beta$, which is the only parameter which cannot be constrained by the parameters of the effective theory at low energies. We chose $\Un{13}=\sin\theta_{13}e^{-i\delta}=0.1i$ and the Majorana phase $\phi_2=1/20$. Interestingly, it is possible to fine-tune the low-energy Majorana phase $\phi_2$ so that $\mathsf{Re}\left(\YY{}\YY{\dagger}\right)_{12}$ is not generated in the leading logarithm approximation and the resulting CP asymmetries are strongly suppressed. We also set the ratio of the vacuum expectation values $\tan\beta_H=\frac{\langle H_u \rangle}{\langle H_d \rangle}=10$. Note that eqs. (\ref{soleps1})--(\ref{soleps2}) provide accuracy about 10--20\% and eqs. (\ref{soleps1ih})--(\ref{soleps2ih}) about a few per cent for vastly different values of $M_N$.

The generated baryon asymmetry depends not only on the CP asymmetries, but also on the strength of the wash-out processes. Since the latter depend very weakly on $\tan\beta_H$, the presence of $y_\tau^2$ in (\ref{soleps1})--(\ref{soleps2}) and (\ref{soleps1ih})--(\ref{soleps2ih}) introduces a strong dependence of leptogenesis on $\tan\beta_H$, i.e. $\epsilon_i \sim \tan^2\beta_H$. In the case of the normal hierarchy, for $M_N\simlt 10^{10}\,\mathrm{GeV}$, one needs $\epsilon_1\sim 10^{-4}$ for successful leptogenesis \cite{Chankowski:2003rr}. Figure \ref{fi1} shows this is a realistic value for rather large $\tan\beta_H\sim 30$. For the inverse hierarchy, the wash-out is roughly twice as big as for the normal hierarchy \cite{Buchmuller:2002rq} and successful leptogenesis can be obtained in this case, as well, by adjusting $\tan\beta_H$ and/or $M_X/M$. The dependence of $\tan\beta_H$ is crucial for successful leptogenesis. In particular, in the non-supersymmetric case $y_\tau^2$ is smaller by a factor $\mathcal{O}(10^2)$ than the value obtained for supersymmetry with $\tan\beta_H=10$ and the mechanism considered here cannot work.

Note that if we adopted a 'texture zero' model, as done in Ref. \cite{GonzalezFelipe:2003fi}, we could easily determine $\beta$ in terms of the low energy parameters. For instance, $\ynu{21}=0$ leads to the prediction:
\beq
\th\beta \approx  \frac{\tan\theta_{13}\sin(\phi_2+\delta)}{\sqrt{\rho}\sin\theta_{12}}+\mathcal{O}\left(\frac{\tan^2\theta_{13}}{\rho}\right)
\eeq
and then the CP violation required for leptogenesis is directly connected to the low energy CP violation.

The minimal see-saw model can be considered a limiting case of the three-right-chiral-neutrino models in which one of the right-chiral neutrinos effectively decouples (either because it is very heavy or because of its vanishingly small Yukawa couplings) \cite{King:1998jw}. To the extent to which the effects of the this right-chiral neutrino can be neglected in the see-saw formula the results considered here apply also to the three neutrino models. There are also models in which the exchange of one of the right-chiral neutrinos gives mass solely to the heaviest light neutrino and the minimal see-saw formalism can be applied to the $\nu_1$, $\nu_2$ sector \cite{Jezabek:2002qu}.

In conclusion, by detailed examination of the RG effects in the right-chiral neutrino sector we obtained the formulae for the CP asymmetries in the supersymmetric minimal see-saw models with degenerate masses of the right-chiral neutrinos. We argued that such models can lead to successful thermal leptogenesis, even though resonant enhancement of the CP asymmetries resulting from small splittings of the masses of the right-chiral neutrinos generated radiatively is partly compensated by small values of $\mathsf{Re}\left(\YY{}\YY{\dagger}\right)$, also generated radiatively. Since, especially for the normal hierarchy of the light neutrino masses, the CP asymmetries depend very weakly on the mass scale of the right-chiral neutrinos, the reheating temperature can be low enough to avoid the gravitino problem. 

\vskip0.3cm
\noindent{\bf Acknowledgments} K.T. would like to thank Marek Je\.zabek for discussion about leptogenesis with degenerate right-chiral neutrinos and  Stefan Pokorski for fruitful discussions and reading the manuscript. Piotr H. Chankowski deserves credit for continuous support. The work was partially supported by the Polish State Committee for Scientific Research grant 2 P03B 129 24 for years 2003-2005.

\end{document}